\documentclass{iau} 
\usepackage{graphicx}

\newcommand{\lsim}{\raisebox{-.4ex}{$\stackrel{<}{\scriptstyle \sim}$}}
\newcommand{\msim}{\raisebox{-.4ex}{$\stackrel{>}{\scriptstyle \sim}$}}

\title[X-ray diagnostics of massive star winds] 
{X-ray diagnostics of massive star winds}

\author[L.\,M. Oskinova, R. Ignace, D.\,P. Huenemoerder]   
{L.\,M. Oskinova$^1$, R. Ignace$^2$, D.\,P. Huenemoerder$^3$}

\affiliation{$^1$Institut f\"ur Physik und Astronomie, 
Universit\"at Potsdam, Germany\\
$^2$Department of Physics and Astronomy, East Tennessee State 
University, TN 37663, USA\\
$^3$Massachusetts Institute of Technology, 
Kavli Institute 
for Astrophysics and Space Research, 70 Vassar St., Cambridge, 
MA 02139, USA}

\pubyear{2016}
\volume{329}  
\jname{The lives and deatch-throes of massive stars}
\editors{J.J. Eldridge, eds.}
\begin{document}

\maketitle

\begin{abstract}
Observations with powerful X-ray telescopes, such as XMM-Newton and  Chandra, 
significantly advance our understanding of massive stars. Nearly 
all early-type stars are X-ray sources. Studies of their X-ray emission provide 
important diagnostics of stellar winds. High-resolution X-ray spectra  
of O-type stars are well explained when stellar wind clumping is taking into 
account, providing further support to a modern picture of stellar winds as 
non-stationary, inhomogeneous outflows. X-ray variability is detected from such 
winds, on time scales likely associated with stellar rotation. 
High-resolution X-ray spectroscopy indicates that the winds of late O-type stars 
are predominantly in a hot phase. Consequently, X-rays provide  
the best observational window to study these winds. X-ray spectroscopy of 
evolved, Wolf-Rayet type, stars allows to probe their powerful metal 
enhanced winds, while the mechanisms responsible for the X-ray emission of these 
stars are not yet understood.  

\end{abstract}

\firstsection 
\section{Introduction}

Stars across the Hertzsprung-Russell diagram (HRD) emit X-rays. The quiescent 
X-ray luminosity of coronal type stars, like our own Sun, is less than one 
per cent of their bolometric luminosity, $L_{\rm x}/L_{\rm bol}\lsim 
10^{-2...-3}$. The X-ray luminosity of massive stars ($M\msim 9\,M_\odot$)
constitutes even smaller fraction of their bolometric luminosity with only 
$L_{\rm x}/L_{\rm bol}\approx 10^{-7}$  \cite[(Pallavicini \etal\ 
1981)]{pal1981}. Yet, despite this small output, X-rays provide invaluable 
diagnostics of massive star winds. 

While it is not yet firmly established which physical processes lead to the
generation of hot plasma emitting X-rays, the observational properties of 
X-ray emission from massive stars are well known 
\cite[(Waldron \& Cassinelli 2007)]{wc2007}. 
The X-ray spectra are dominated by strong emission lines, and are consistent 
with 
being produced by fast expanding hot optically thin plasma. Spectral diagnostics 
indicate that the hot plasma observed in the X-ray band is permeated with the 
cool 
stellar wind best observed at UV wavelengths. 

The classical diagnostics of OB star winds is provided by the modeling of 
their UV spectra. In stars with strong winds, the resonance lines of metal ions 
usually exhibit P Cygni type profiles. Fitting these lines with a  model, e.g.\ 
based on the Sobolev approximation \cite[(Hamann 1981)]{hamann1981}, 
provides information on the product of mass-loss 
rate, $\dot{M}$, and the ionization fraction. Only when the ionization fraction 
is well constrained, one can reliably estimate mass-loss rates. It was, 
however,  
noticed early that X-ray photons in stellar wind serve as additional source 
of ionization of metal ions, chiefly via Auger process  
\cite[(Cassinelli \& Olson 1979)]{cas1979}. This effect 
is often referred to as ``superionization''. Therefore, to measure mass-loss 
rates, especially from stars with weaker winds, the calculations of ionization 
balance shall account for the influence of X-rays. \cite[Oskinova \etal\ 
(2011)]{osk2011} studied the UV spectra of a sample of B-type stars. The 
observed 
spectra were modeled with the non-LTE stellar atmosphere model PoWR that 
accounts 
for the X-ray radiation. From comparison between models and observations, it 
was found that the winds of the stars in our sample are quite weak. The 
wind velocities 
do not exceed the escape velocity. The X-rays strongly affect the ionization structure 
of these winds. But this effect does not significantly reduce the total 
radiative 
acceleration. Even when X-rays are accounted for, there is still sufficient radiative 
acceleration to drive a stronger mass-loss than empirically inferred from 
fitting the UV resonance lines. These findings are in line with conclusions 
reached by \cite[Prinja (1989)]{pr1989}. 

\section{The ``weak wind problem''}
\label{sec:ww}

Weak winds are also encountered in late O-type dwarfs \cite[(Marcolino \etal\ 
(2009)]{mar2009}. 
\cite[Lucy (2012)]{lucy2012} suggested a model explaining the low 
mass-loss rates empirically derived for OB dwarfs. He proposed that 
shock-heating increases the  temperature of the gas, leading to a 
temperatures of a few MK. The single component flow coasts to high
velocities as a pure coronal wind. The model predicts that the bulk 
of stellar wind is hot. Only some small fraction of the wind remains 
cold and is embedded in the hot wind.    

To study the wind of a typical O9V star, we  analyzed 
high-resolution X-ray spectra of $\mu$\,Col  \cite[(Huenemoerder 
\etal\ 2012)]{huen2012}. The analysis of the spectra did not reveal  
any significant traces of absorption of X-rays by the cool wind component, 
in agreement with  Lucy's prediction. The analyses of line 
ratios  of He-like ions revealed that the hot matter is  present already 
in the inner wind. On the other hand, the  X-ray emission lines are
broadened up to terminal wind velocity, $v_\infty$.  The shape of the 
X-ray emission 
line profiles can be well described as originating from a hot plasma expanding 
with a usual $\beta$-velocity law. Moreover, the emission measure of the hot 
wind 
component is quite large, exceeding that of the cool wind. 
Considered together, these findings point out that the winds of O-dwarfs 
likely are
predominantly  in the hot phase, while the cool gas seen in the  optical
and UV  constitutes only a minor wind fraction.  Hence, the best 
observational window for studies of OB dwarfs is provided by X-rays.  

\section{X-ray pulsations in massive stars}
\label{sect:bcep}

Monitoring X-ray observations allow to study wind variability. Observations 
of pulsating B-type stars  provide excellent means to investigate the links 
between stellar wind and stellar interior. Young B0-B2 type stars 
that are still burning hydrogen in their cores oscillate with periods 
of a few hours and are known as $\beta$\,Cephei-type 
variables \cite[(Dziembowski \& Pamiatnykh 1993)]{dz1993}.

Recently, it has been shown that the X-ray emission from the  magnetic 
star {$\xi^1$\,CMa} pulsates in phase  with the  optical 
\cite[(Oskinova \etal\ 2014)]{osk2014}. 
Strong phase dependent variability was also detected in the high-resolution 
X-ray spectrum. The variability in N\,{\sc vii}\,$\lambda 24.8$\,\AA\ line 
was attributed to changes in wind ionization structure with 
stellar pulsational phase. Spectral diagnostics revealed that the hot 
X-ray emitting plasma is located very close to the stellar photosphere.
The physical mechanism causing the X-ray pulsations is not yet 
known, but one may speculate that surface magnetic fields may be involved in 
coupling the wind to subphotospheric layers.  Coming X-ray 
and UV observations of a representative sample of $\beta$\,Cephei variables 
will shed more light on this question.     

\section{X-ray variability of OB supergiants}
\label{sect:rot}

Coherent and periodic  variability  is commonly observed in 
the UV lines of OB supergiants \cite[(Kaper \etal\ 1999)]{kap1999}. 
This variability is likely explained  by the existence of corotating interaction
regions  (CIRs) in stellar winds \cite[(Mullan 1984, Hamann \etal\ 
2001)]{mullan1984, hamann2001}. \cite[Cranmer \& Owocki (1996)]{Cranmer1996}   
showed that
CIRs could result from bright stellar spots.  \cite[Ramiaramanantsoa 
\etal\ (2014)]{Ram2014} detected
corotating  bright spots on $\xi$\,Per\ and suggested  that they
are  generated via a breakout of a global magnetic field generated by 
subsurface convection. The CIRs may also be triggered by the
(non)radial  pulsations  of the stellar surface 
\cite[(Lobel \& Blomme 2008)]{Lobel2008}.

\cite[Oskinova \etal\ (2001)]{osk2001} found the X-ray light-curve of the 
O-type dwarf 
$\zeta$\,Oph being modulated on the rotation time scale, with a period 
similar to the one observed in UV lines.  Similar conclusions on X-ray 
variability in O stars were reached by \cite[Massa \etal\ (2014)]{Massa2014} 
who analyzed X-ray observations of $\xi$\,Per  (O7.5III) obtained with the {\em 
Chandra}
X-ray telescope and contemporaneous  H$\alpha$ observations. The 
X-ray flux was found to vary by $\sim 15$\%, but  not in phase with 
the H$\alpha$ variability. The observations were not  long enough
to establish periodicity.

Among the O-stars best monitored in X-rays  is the O4 supergiant $\zeta$\,Pup. 
The {\em XMM-Newton} X-ray telescope observed it from time to time during a 
decade. Analysis of the obtained light curves revealed 
variations with an amplitude of $\sim 15$\%\ on a time scale longer than 
1\,d, while no coherent periodicity was detected 
\cite[(Naz\'e \etal\ 2013)]{naze2013}. Interestingly, 
\cite[Howarth \& Stevens (2014)]{how2014} reported a period of $P=1.78$\,d  
in optical photometry of this star, which they attributed to stellar 
pulsations. 

In another O-type supergiant, $\lambda$ Cep (O6.5I(n)), the X-ray flux also  
varies by $\sim 10$\,\%\ on  timescales of days, possibly modulated with 
the same period as the H$\alpha$ emission \cite[(Rauw \etal\ 2015)]{rauw2015}.  
The analysis of archival {\em XMM-Newton} observations of $\zeta$ Ori (O7I) 
shows that the X-ray variability of this star has similar properties  to
that of $\lambda$ Cep and $\zeta$ Pup. X-ray variability with analogous 
character was found in $\delta$ Ori (O9.5II+B1V) from {\em Chandra} 
observations, where periodic fluctuations (not associated  with
its binary period) were identified \cite[(Nichols \etal\ 2015)]{Nichols2015}. 
Even Wolf-Rayet (WR) 
stars with  very strong winds show modulations in their X-ray  emission 
on the rotation timescale  \cite[(Ignace \etal\ 2013)]{Ignace2013}. 
Summarizing, the evidence for X-ray variability on rotation 
time-scale 
is accumulating. This points to an association between X-ray emission, large 
scale structures in the stellar wind, and stellar spots.

\section{X-ray emission lines in spectra of OB supergiants}
\label{sect:mdot}

Over the last decade, the analysis of emission lines observed in  
X-ray spectra  has become a common tool for stellar wind studies. 
The profile of an emission line originating from an optically thin 
expanding shell was derived by \cite[Macfarlane \etal\ (1991)]{mac1991}. 
It was shown that continuum absorption in the stellar 
wind leads to characteristically blue-shifted and skewed line profiles, 
and suggested that the ``skewness'' of a line could be used to 
estimate the wind column density. This model was further 
extended and applied to X-ray emission line profiles \cite[(Waldron \&
Cassinelli 2001, Owocki \& Cohen 2001, Ignace 2001)]{WC2001, OC2001, RI2001}. 
Stellar wind clumping was included in the modeling by 
\cite[Feldmeier \etal\ (2003)]{Feld2003}. They showed 
shown that clumping reduces the effective wind opacity and affects the shape of 
the emission line profiles. In general, lines emerging from clumped winds
are less skewed than those produced in homogeneous winds even when the 
mass-loss rates are the same \cite[(Oskinova \etal\ 2006)]{osk2006}. 

\begin{figure}
\centering
\includegraphics[width=6cm]{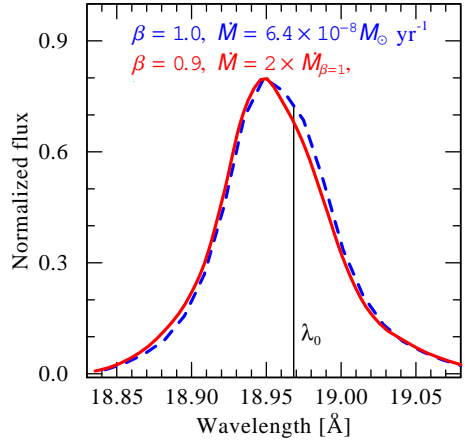}
\caption{Illustration of the degeneracy of the X-ray emission line 
profiles with respect to the velocity field. The vertical line indicates 
the central wavelength of the O\,{\sc viii} L$\alpha$ line. Two model 
lines are shown, each of them  computed using the same ``universal'' 
mass-absorption coefficient $\kappa$ and stellar parameters as suggested by 
\cite[Cohen \etal\ (2014)]{coh2014}. The blue-dotted profile is based on a 
smooth wind model with $\beta=1$, onset of the X-ray emission at 
$R_0=1.33\,R_\ast$, and $\dot{M}=6.3\times 10^{-8} M_\odot$\,yr$^{-1}$. The 
red-solid line is a smooth wind model, but now for $\beta=0.9$, 
$R_0=1.26\,R_\ast$, and $\dot{M}=1.2\times 10^{-7} M_\odot$\,yr$^{-1}$, i.e. 
twice as high.}
  \label{fig:betam}
\end{figure}

Besides continuum absorption and wind clumping, the shape of an X-ray 
emission line also depends on the assumed velocity field (as illustrated in 
Figure\,\ref{fig:betam}), wind geometry, abundances, ionization balance, hot 
plasma distribution, onset of X-ray emission and resonance scattering in the 
hot plasma. This incomplete list shows that, contrary to claims in the 
literature, it is hardly possible to derive just one parameter, such as 
mass-loss rate, from fitting the shapes of X-ray emission lines by simplistic 
models.

On the other hand, X-ray spectroscopy complemented by the analysis of 
UV and optical spectra using non-LTE stellar atmosphere models is 
a valuable tool to constrain stellar wind properties \cite[(Herv\'e 
\etal\ 2013,  Puebla \etal\ 2016)]{herve2013, pu2016}. E.g.,\ \cite[Shenar 
\etal\ (2015)]{Shenar2015} applied PoWR non-LTE stellar atmospheres for the 
analysis  of X-ray, optical, and UV spectra of the O star $\delta$\,Ori. 
It was possible to reproduce the UV and optical spectra as well as the X-ray 
emission lines consistently, and thus derive realistic wind parameters.

\section{X-ray emission from blue hypergiants and Wolf-Rayet stars}
\label{sect:mdot}

Among those stars with the most powerful stellar winds are the blue hypergiants 
and the WR stars. X-ray spectroscopy provides interesting insights to their 
stellar winds.  

Recently, we analyzed the X-ray spectrum of one of the most massive and 
luminous 
stars in the Milky Way, Cyg\,OB2-12 ((B3Ia$^+$), obtained with the {\em 
Chandra} X-ray telescope. The analysis was complemented by modeling, 
using the PoWR atmosphere model  (Oskinova \etal, in prep.). 

It was shown that the X-ray spectrum of Cyg\,OB2-12 is produced in a hot plasma 
with temperatures in excess of 15\,MK. Given that the stellar wind of this star 
is rather slow, $v_\infty\approx 400$\,km\,s$^{-1}$, it is difficult to explain 
high temperature by intrinsic wind shocks. From the ratio of fluxes in 
forbidden and intercombination lines in He-like ions followed that 
that the X-ray emitting plasma has quite a high density. Furthermore, the 
broadening of X-ray emission lines exceeds the one due to the stellar wind 
velocity. Taken together, these facts are best explained by a colliding winds 
scenario. Further support to the binary hypothesis is provided by the recent 
identification of a companion in the Cyg\,OB2-12 system \cite[(Maryeva \etal\ 
2016, and ref. therein)]{mar2016}. 


The low-resolution X-ray spectra of  WR stars are well described by 
thermal plasmas  with  temperatures  between 1\,MK up to 
50\,MK \cite[(Ignace \etal\ 2003, and ref. therein)]{Ignace2003}.  
High-resolution X-ray spectra are available so far only for one single  WR star 
- EZ CMa (WR\,6) \cite[(Oskinova \etal\ 2012)]{osk2012}. Their analysis shows 
that hot plasma exists far out in the wind. The X-ray emission lines are broad 
and strongly skewed (just as predicted by the \cite[Macfarlane \etal\ (1991) 
and  Ignace (2001) models)]{mac1991, RI2001} being consistent with plasma 
expanding with constant velocity. The abundances derived from the X-ray spectra 
are strongly non-solar, in accordance with the advanced evolutionary 
state of WR stars \cite[(Huenemoerder \etal\ 2015)]{Hue2015}. How X-rays are 
generated in winds of single WR stars is not yet understood.  It is possible  
that, like in O type stars, X-ray generation is an intrinsic ingredient of the  
stellar wind driving  \cite[(Gayley 2016)]{gay2016}. 

To conclude, X-ray observations provide important diagnostics of massive stars 
and their winds at nearly all evolutionary stages.

\end{document}